\documentclass[graybox]{svmult}

\usepackage{mathptmx}       
\usepackage{helvet}         
\usepackage{courier}        
\usepackage{type1cm}        
\usepackage{makeidx}         
\usepackage{graphicx}        
\usepackage{multicol}        
\usepackage[bottom]{footmisc}

\makeindex             

\begin{document}

\title*{Singular Spectrum Analysis for astronomical time series:
constructing a parsimonious hypothesis test}
\titlerunning{Singular Spectrum Analysis for astronomical time series} 
\author{G. Greco 
\and D. Kondrashov 
\and S. Kobayashi 
\and M. Ghil
\and M. Branchesi
\and C. Guidorzi 
\and G. Stratta
\and M. Ciszak    
\and        F. Marino
\and        A. Ortolan
}
\authorrunning{Singular Spectrum Analysis for astronomical time series} 
\institute{G. Greco, M. Branchesi, G. Stratta  \at (1) 
Universit\`a degli Studi di Urbino ``Carlo Bo'' -- DiSBeF -- I-61029 Urbino (PU), Italy;
(2) INFN, Sezione di Firenze, Via Sansone 1, I-50019 Sesto Fiorentino (FI), Italy
\and D. Kondrashov and M. Ghil(*) \at  University of California --AOS and IGPP -- Los Angeles, CA 90095-1565, USA
and (*) Ecole Normale Sup\'erieure -- CNRS and IPSL -- F-75231 Paris Cedex 05, France
\and S. Kobayashi \at ARI - Liverpool John Moores University,146 Brownlow Hill, Liverpool, L3 5RF, UK 
\and C. Guidorzi  \at University of Ferrara, Department of Physics and Earth Sciences, I-44122 Ferrara, Italy
\and M. Ciszak and F. Marino \at CNR-Istituto Nazionale di Ottica, L.go E. Fermi 6, I-50125 Firenze, Italy
\and A. Ortolan \at INFN, Laboratori Nazionali di Legnaro, I-35020 Legnaro (PD), Italy 
}
%
%

\maketitle


\abstract{
We present a data-adaptive spectral
method - Monte Carlo Singular Spectrum Analysis (MC-SSA)  - and its modification to tackle astrophysical problems. Through numerical simulations we show the ability of the MC-SSA in dealing with $1/f^{\beta}$ power-law noise  affected by photon counting statistics. 
Such noise process is simulated
by a first-order autoregressive, AR(1) process corrupted by intrinsic Poisson noise.
In doing so, we statistically estimate a basic stochastic variation of the source and the corresponding fluctuations due to the quantum nature of light.
In addition,  MC-SSA test retains its effectiveness
even when a significant percentage of the signal falls below a certain level of detection,
$e.g.,$ caused by the instrument sensitivity.
The parsimonious approach presented here may be broadly applied, from the search for extrasolar planets to the extraction of low-intensity coherent phenomena probably hidden in high energy transients.
}

\section{Colored noise and MC-SSA}
\label{sec:1}
$1/f^{\beta}$ power-law noise is known to be highly relevant in several astrophysical systems
\cite{pont06,vau10,greco2011}.
It includes the well-known white noise ($\beta$ = 0), pink noise ($\beta$ = 1) and red or Brownian noise ($\beta$ = 2).
Such kind of noises
 are  very often not tied to instrumental disturbance but rather they are 
property of the observed sources that emit radiation varying in a stochastic manner.
In the context of our work, 
it is sufficient to qualitatively 
characterize $1/f^{\beta}$ to generate a parsimonious hypothesis test in MC-SSA (see below).  
Subsequently, the dynamical origin of noise fluctuation or the extra-noise variance 
-- according to acceptance or rejection of our null hypothesis --
has to be determined by theoretical analysis.

Singular Spectrum Analysis (SSA) is an effective, data-adaptive and non-parametric method for the decomposition of  
a time series into a  well-defined set of independent and interpretable components that include a non-linear trend, anharmonic,  amplitude-modulated oscillations, and noise \cite{VauGh89}.
In its later developments,
the Monte Carlo approach to signal-to-noise separation introduced by  \cite{allen1996} has become known as Monte Carlo SSA (MC-SSA) \cite{ghil2002}. 
The distinction between what is  $noise$ and what is  $signal$ is made through the measure 
of the resemblance of a given noise surrogate to the original data via eigendecomposition of the time-lagged covariance matrix.
For this study the noise surrogates set is generated by using an AR(1) process.
In practice, the AR(1) coefficients  are estimated from the time-series under 
 consideration by using a maximum-likelihood criterion. 
Subsequently, these AR(1) noise surrogates are corrupted by Poisson noise to mimic the effect of a  photon counting detector. 


\section{Simulated $1/f^\beta$ series test}
\label{subsec:2}
The procedure has been tested on a large sample of artificial time series of arbitrary colored noises.
As a test of validity, we show several discrete colored noise vectors of
length $N$ = 5000, with a  power-law 
distribution of  slope $\beta$ ranging from 1 to 2 in steps of two.
To obtain sequences of colored noises we use the Matlab library CNOISE\footnote{http://people.sc.fsu.edu/~jburkardt/m$\_$src/cnoise/cnoise.html}.
A threshold of signal identification  is incorporated by subtracting a mean value
$\sim$ 35$\%$ of the maximum peak-flux, $C_{max}$.
In this way, we take into account episodes of emission during which the emission count rate drops to the background level. 
Finally, the time series are corrupted by Poisson noise to simulate the effect of the  shot noise
based on Poisson photon statistics. 
The resulting test series, 
are shown in the supplementary video\footnote{https://vimeo.com/120699083}.
The  signal to noise ratio (SNR) of the artificial series test increases as the simulations run.

\section{Conclusion and future work}
We present a modified data-adaptive spectral method -- Monte Carlo Singular Spectrum Analysis (MC-SSA) --
in which a parsimonious (dynamic and instrumental) noise-model $H_{0}$ is adopted for astrophysical applications, i.e. $H_{0}$ = \{AR(1) + Poisson noise\}. AR(1) noise takes into account the long-term
variability in the  power-law slope -$\beta$ of a light-curve's spectral density, 
and the Poisson noise considers the short-term variability imposed by the quantum nature of light.
Our simulations show a remarkable effectiveness of the model for colored noises with $\beta$ values between 1.5 and 2
and with $C_{max}$ $>$ 1000 counts; otherwise one should re-bin the light-curve to improve the  signal-to-noise ratio of the source.

All analyses were performed by using SSA-MTM Toolkit freeware that has been 
developed by the Theoretical Climate Dynamics Group at UCLA\footnote{http://www.atmos.ucla.edu/tcd/ssa/}. 
To obtain presented results we relied on an advanced option in SSA-MTM Toolkit for MC-SSA that allows to read in ensemble of surrogates created off-line and corresponding to an arbitrary noise null-hypothesis $H_{0}$. 
In this regard, MC-SSA allows to iteratively refine the working hypothesis 
and adapt the null-hypothesis $H_{0}$ for various theoretical scenarios.
We plan to extend further SSA-MTM Toolkit for astrophysical applications by directly including the algorithm for parsimonious $H_{0}$ in the freeware. 
Current and future work will focus on SSA as a promising technique for analysing complex spatio-temporal structures to detect possible periodicities and other statistical regularities due to various astrophysical processes.
 

\begin{acknowledgement}
MB, GG and GS  acknowledge the financial support  of the Italian Ministry of Education, University and Research (MIUR) through grant FIRB 2012 RBFR12PM1F.
CG acknowledges the PRIN MIUR project on "Gamma Ray Bursts: from progenitors to physics
of the prompt emission process" (Prot. 2009 ERC3HT).  
MG and DK received support from the U.S. National Science Foundation (grant DMS-1049253) and from the U.S. Office of Naval Research (MURI grant N00014-12-1-0911). 
\end{acknowledgement}

%

\end{document}